\documentclass{article}
\usepackage{graphicx}
\begin{document}
%+Titl
\title{ {\huge  Mona Lisa}\\ the
stochastic view and fractality in color space}
\author{Pouria Pedram$^{a,b}$\thanks{pouria.pedram@gmail.com}, G.~R.~Jafari$^a$\thanks{gjafari@gmail.com}
\\ {\small $^a$ Department of Physics, Shahid Beheshti University, Evin,
Tehran 19839, Iran}\\{\small $^b$ Plasma Physics Research Center, Science and Research Campus,}\\{\small Islamic Azad University, Tehran, P.O.Box: 14665-678, Iran}
}
\date{\today}
\maketitle \baselineskip 24pt
%-Title
\begin{abstract}
A painting consists of objects which are arranged in specific ways.
The art of painting is drawing the objects, which can be considered
as known trends, in an expressive manner. Detrended methods are
suitable for characterizing the artistic works of the painter by
eliminating trends. It means that we study the paintings, regardless
of its apparent purpose, as a stochastic process. We apply
multifractal detrended fluctuation analysis to characterize the
statistical properties of Mona Lisa, as an instance, to exhibit the
fractality of the painting. Our results show that Mona Lisa is long
range correlated and almost behaves similar in various scales.
\end{abstract}

\textit{Pacs}:{02.50.Fz, 05.45.Tp}

\section{Introduction}\label{sec1}
Art is a collection of objects created with the intention of
transmitting emotions and/or ideas. An object can be characterized
by intentions of its creator, regardless of its apparent purpose. In
this sense, art is described as a deliberate process of arrangement
by an agent. Art stimulates an individual's thoughts, emotions,
beliefs, or ideas through the senses. It is also an expression of an
idea and it can take many different forms and serve many different
purposes. Mathematics and art are examples of human motivations to
understand reality
\cite{taylor,Kandinsky,bach,taylor2,mozart,music}. The connection
between mathematics and art goes back to thousands of years.
Patterns, symmetries, proportions and transformations are
fundamental concepts common to both disciplines. One of the
connections between mathematics and art is that some known as
artists have needed to develop or use mathematical thinking to carry
out their artistic insight. Thus we are forced to use interpretative
techniques in order to search for them. In particular, there are
several obvious relations between art and mathematics such as music
\cite{bach,mozart,music,music2,music3,music4} and painting
\cite{taylor,taylor2,taylor3,taylor4,Kandinsky,painting,painting2,painting3}.

Painting has three principal parts: drawing, proportion and
coloring. Drawing is outlines and contours contained in a painting.
Proportion is these outlines and contours positioned in proportion
in their places. Coloring is giving colors to things. One can
obviously use mathematical geometry to analyze paintings, figurative
or abstract, in terms of shapes such as points and lines, circles
and triangles. One may think of the application of the perspective
theory to figurative painting, or the use of some concepts like
fractals for comprehension of abstract paintings. An important
example is the analysis of Jackson Pollock's drip paintings in terms
of fractal geometry \cite{taylor}.

Here, we are interested to study the fractal nature of painting from
another point of view which in general, could be applicable to
non-stationary series \cite{bach}. Among various paintings we focus
on characterizing the complexity of color signal of ``Mona Lisa''
through computation of signal parameters and scaling exponents,
which quantifies correlation exponents and multifractality of the
signal.

\begin{figure}
\begin{center}
\includegraphics[width=6cm]{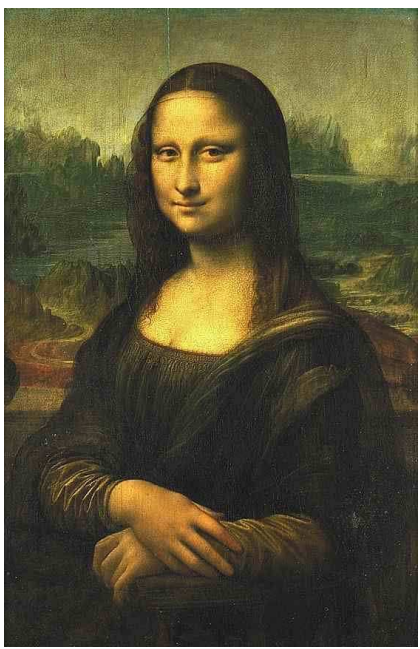}
\end{center}
\caption{\label{fig1} Mona Lisa: oil painting on poplar wood by Leonardo da
Vinci.}
\end{figure}

Mona Lisa, or La Gioconda (La Joconde), is a 16th century oil
painting on poplar wood by Leonardo da Vinci, and is the most famous
painting in the world. This painting is a half-length portrait which
depicts a woman whose gaze meets the viewer's with an enigmatic
expression (Fig.~\ref{fig1}). Because of non-stationary nature of
color signal series, and due to finiteness of available data sample,
we should apply methods which are insensitive to non-stationarities,
like trends. In order to separate trends from correlations we need
to eliminate trends in our color data. Several methods are used
effectively for this purpose: detrended fluctuation analysis (DFA)
\cite{Peng94}, rescaled range analysis (R/S) \cite{hurst65} and
wavelet techniques (WT) \cite{wtmm}. However, using
increment series is customary to make a stationary series from a non-stationary
ones \cite{return}.

%@@@@@@@@@@@@@@@@@@@@@@@@@@@@@@@@
%@@@@@@@@@@@@@@@@@@@@@@@@@@@@@@@@

\begin{figure}
\centering
\includegraphics[width=4cm]{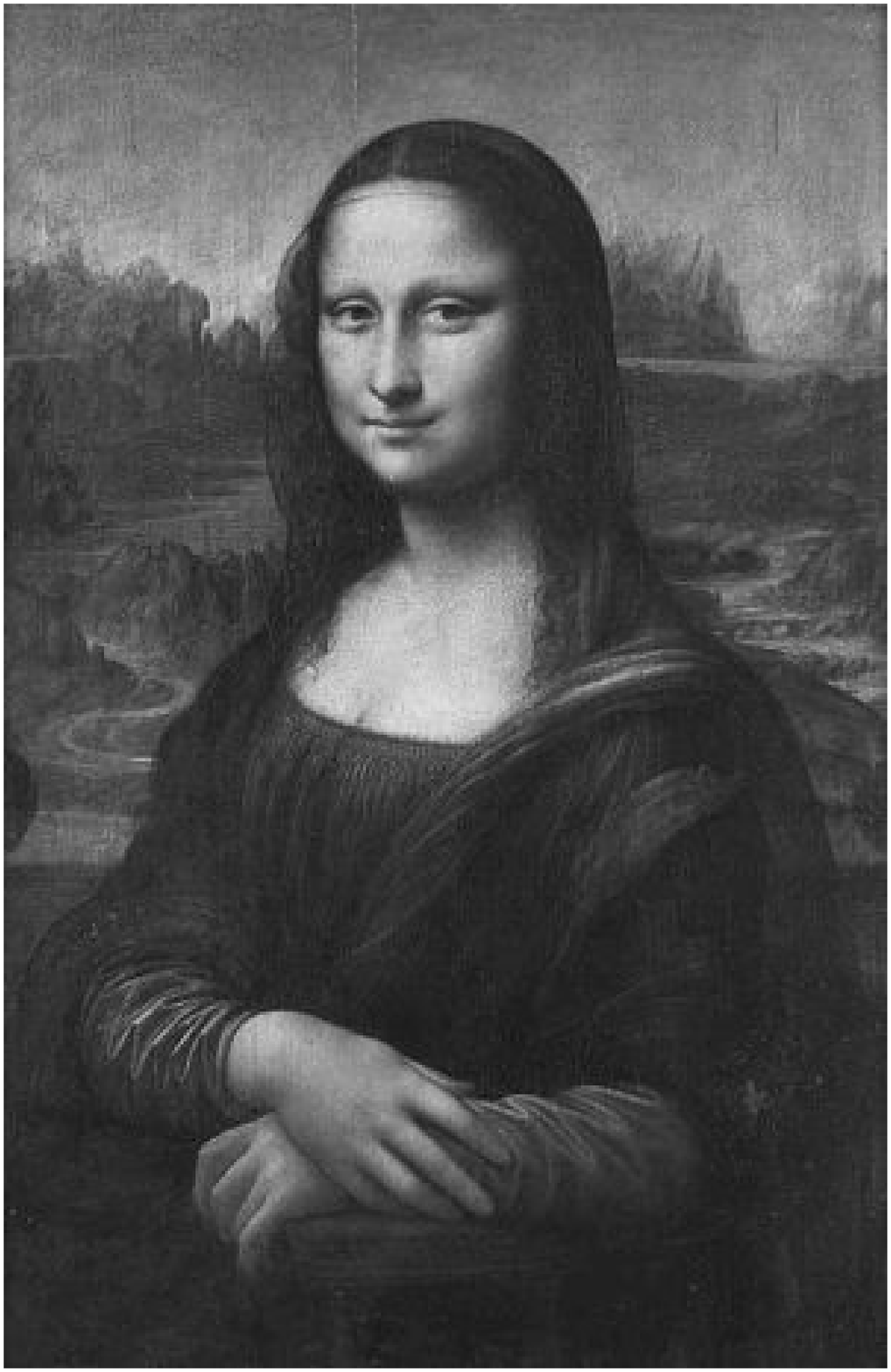}
\begin{tabular}{ccccc}
\includegraphics[width=4cm]{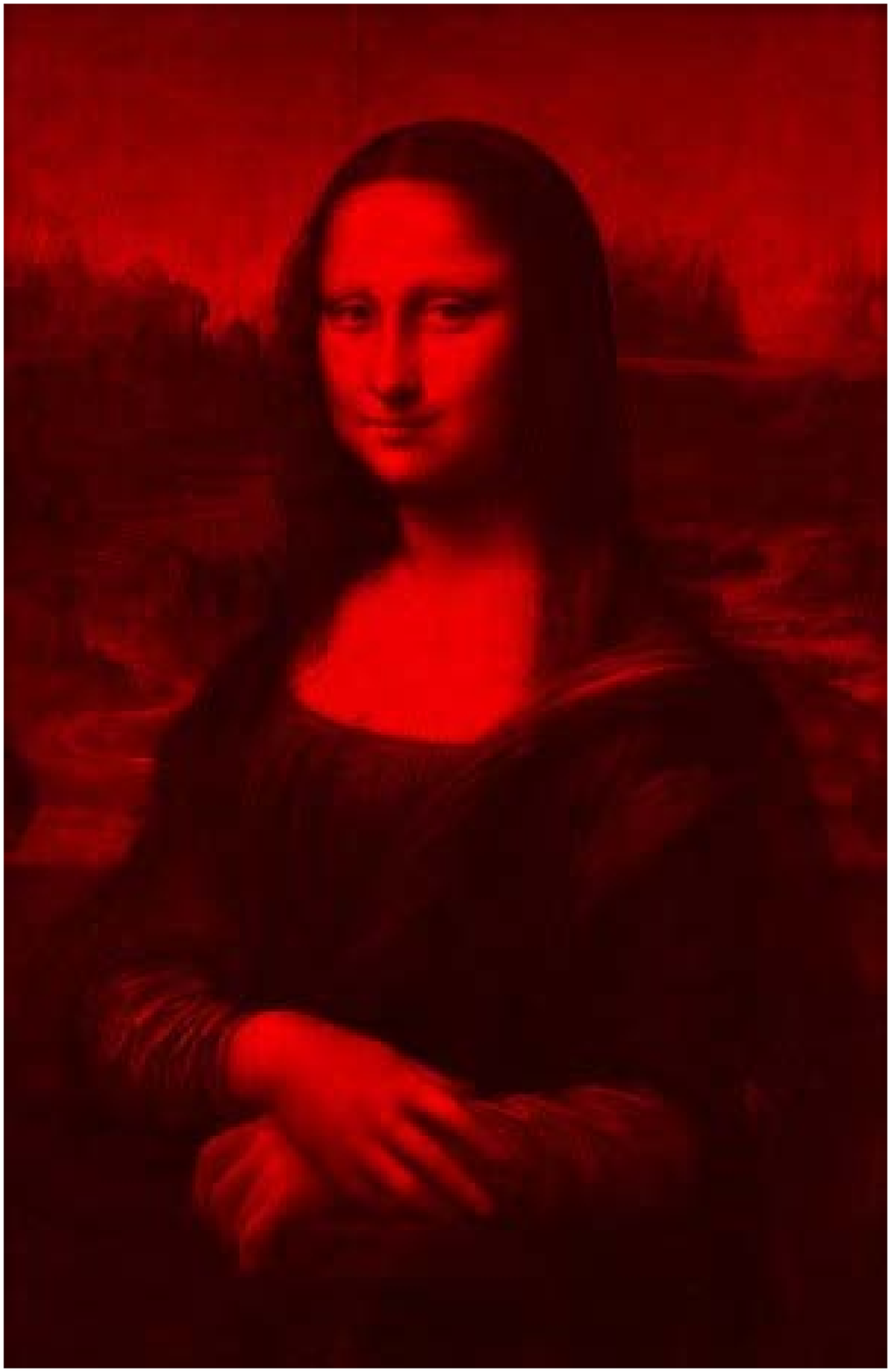}
 &\hspace{1.cm}&
\includegraphics[width=4cm]{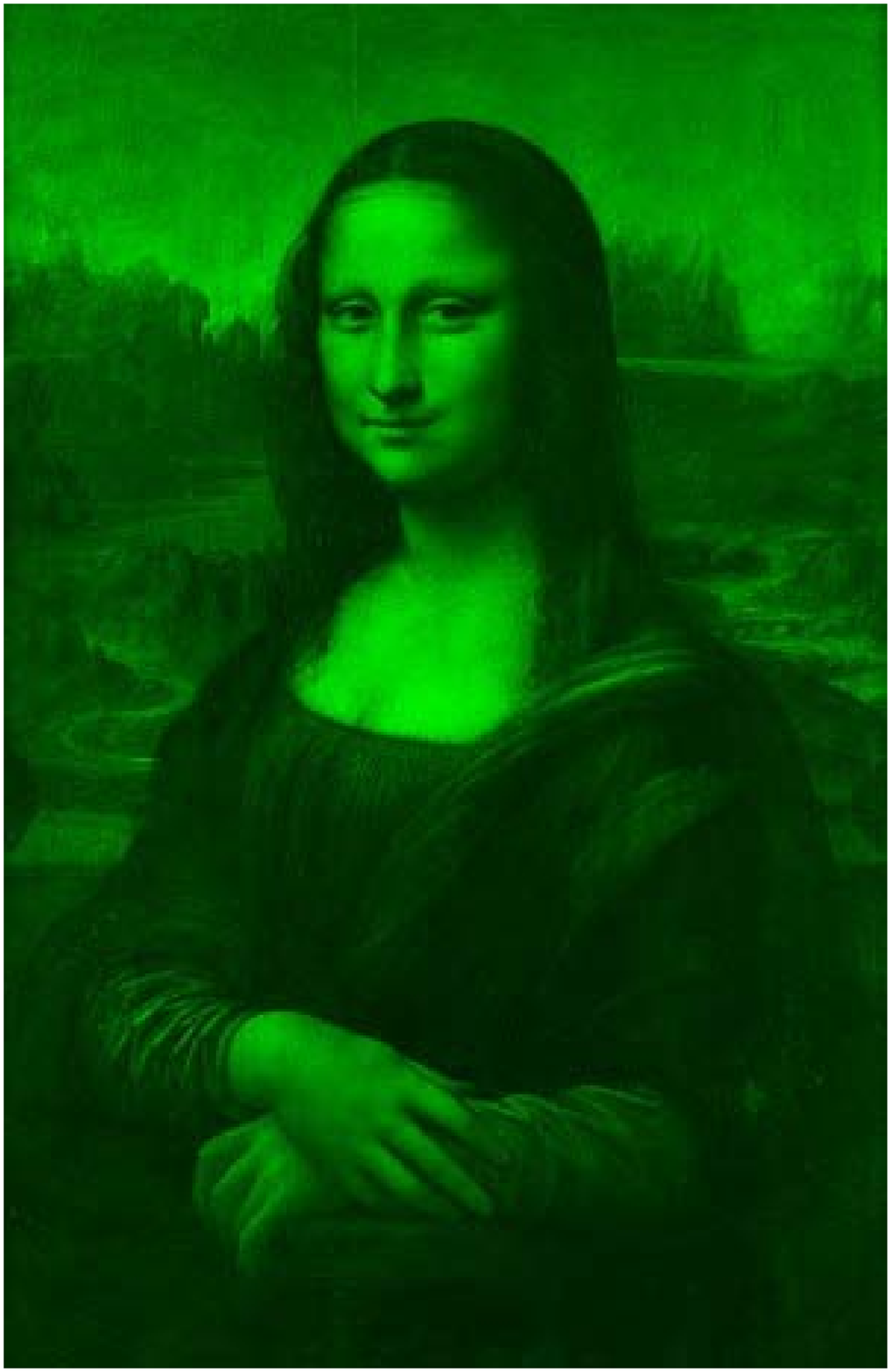}
 &\hspace{1.cm}&
\includegraphics[width=4cm]{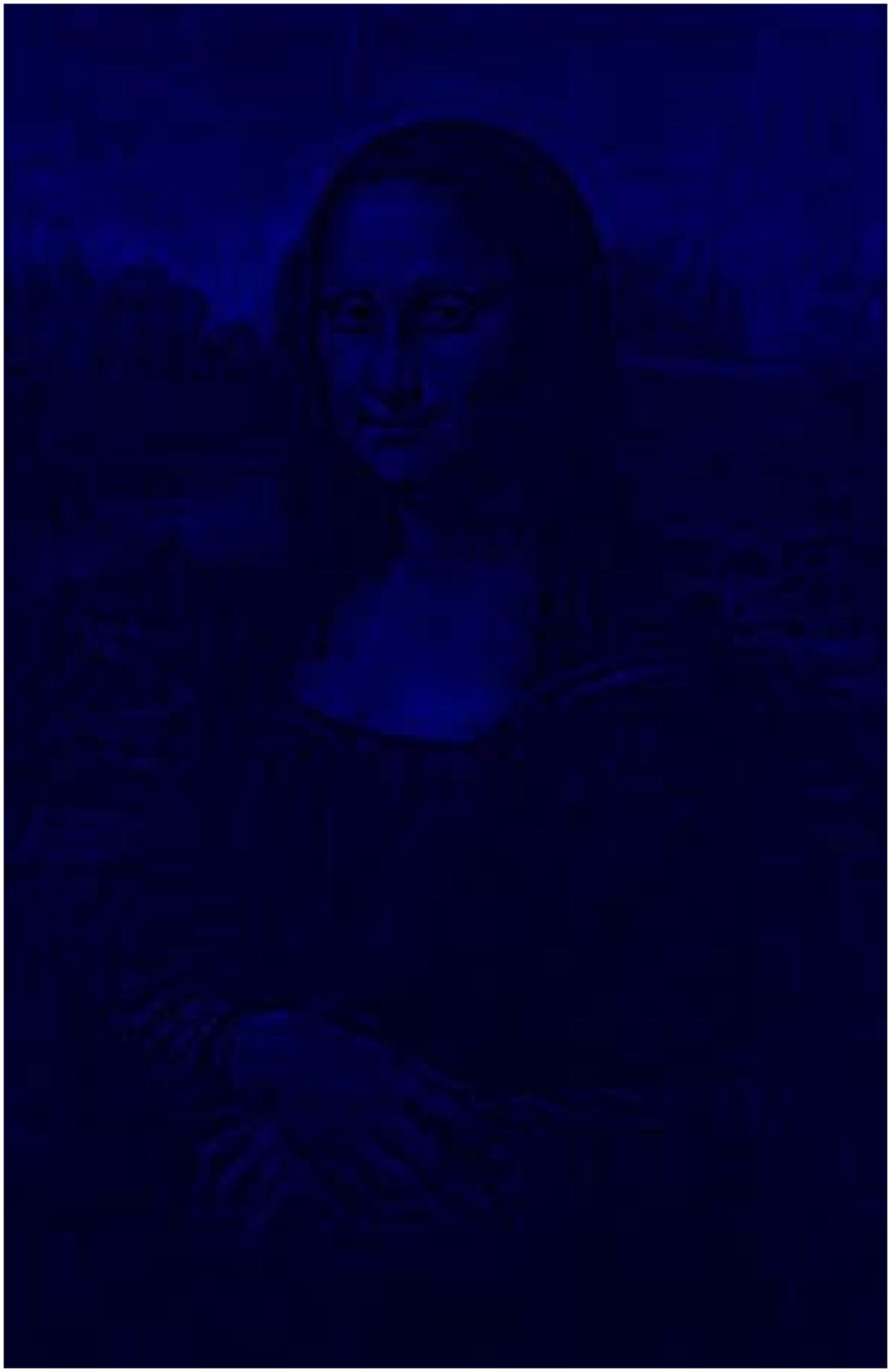}
\end{tabular}
  \caption{\label{fig2} Gray, red, green and blue style pictures (up to down respectively) of Mona Lisa}
\end{figure}

\begin{figure}
\centering
\includegraphics[width=12cm]{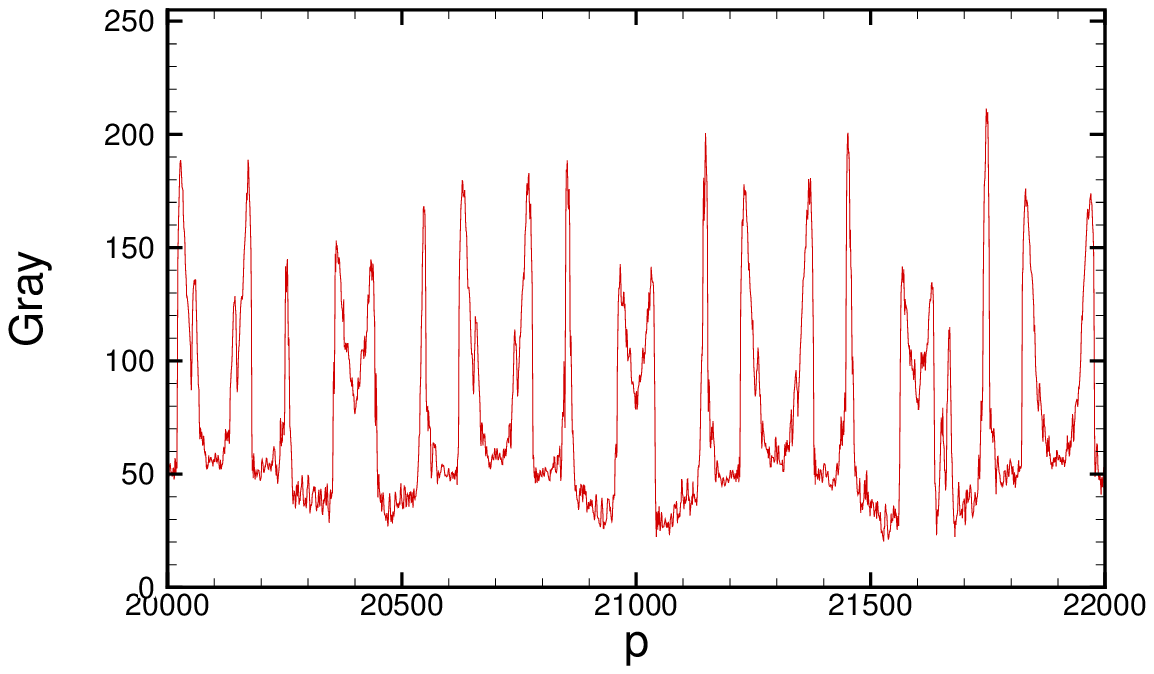}
\\
\includegraphics[width=12cm]{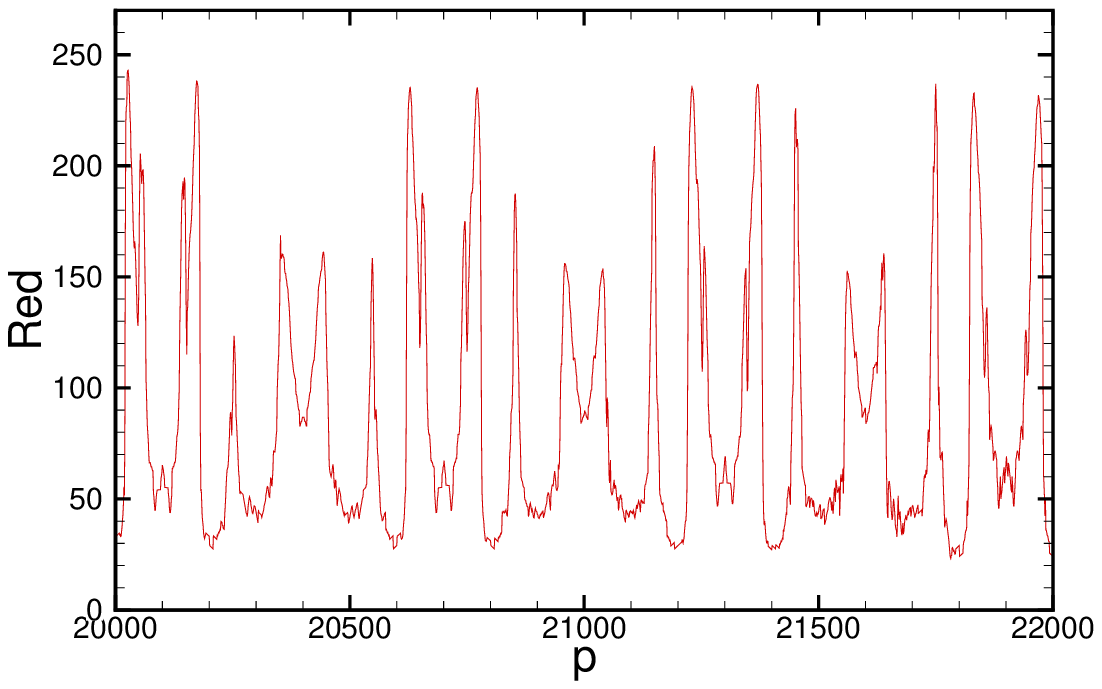}
  \caption{\label{fig3} Typical color value of color series for gray (Up) and red (Down) style.}
\end{figure}

We use MF-DFA method for analysis and eliminating trends from data
set. This method is the modified version of DFA method to detect
multifractal properties of time series. DFA method introduced by
Peng \textit{et al.} \cite{Peng94} has become a widely used
technique for the determination of monofractal scaling properties
and the detection of long-range correlations in noisy,
non-stationary time series \cite{murad,physa,kunhu,kunhu1}. It has
successfully been applied to diverse fields
\cite{Peng94,dns,herz,Peng95,PRL00,economics,bach,ausloos1,ausloos3,ausloos6,sadegh}.
One reason to employ DFA method is to avoid spurious detection of
correlations that are artefacts of non-stationarity series. The
focus of the present paper is on the fractality nature of color
series obtained from Mona Lisa. To construct the series, we
calculate the standard color values of each pixel of the picture
successively for each row from up to down, continuously. In
particular, Fig.~\ref{fig1} shows Mona Lisa and Fig.~\ref{fig2}
shows its gray, red, green and blue color styles for a
$202\times300$ pixels sample. The color fluctuation graphs of gray
and red styles, are also depicted in Fig.~\ref{fig3}.

The paper is organized as follows: In Sec.~\ref{sec2}, we describe
MF-DFA methods in detail and show that scaling exponents determined
by MF-DFA method are identical to those obtained by standard
multifractal formalism based on partition functions. In
Sec.~\ref{sec3}, in analysis of color series of Mona Lisa we also
examine the multifractality in color data. Section \ref{sec4} closes
with a conclusion.

\section{Multifractal Detrended Fluctuation Analysis}\label{sec2}

The simplest type of multifractal analysis is based upon standard
partition function multifractal formalism, which has been developed
for multifractal characterization of normalized, stationary
measurements \cite{feder88,barabasi,peitgen,bacry01}. Unfortunately,
this standard formalism does not give us correct results for
non-stationary time series that are affected by trends or those
which cannot be normalized. MF-DFA is based on identification of
scaling of $q$th-order moment depending on signal length, and this
is a generalization of standard DFA method in which $q=2$. Moreover
one should find correct scaling behavior of fluctuations, from
experimental data which are often affected by non-stationary
sources, like trends. These have to be well distinguished from
intrinsic fluctuations of the system. In addition, often in
collected data we do not know the reasons, or even worse the scales,
for underlying trends, and also available record data is usually
small. So, for a reliable detection of correlations, it is essential
to distinguish trends for intrinsic fluctuations from collected
data. Hurst rescaled-range analysis \cite{hurst65} and other
non-detrending methods work well when records are long and do not
involve trends, otherwise they might give wrong results. DFA is a
well established method for determining scaling behavior of noisy
data where the data include trends and their origin and shape are
unknown \cite{Peng94,Peng95,fano,allan,dns}.

Modified multifractal DFA procedure consists of five steps. The
first three steps are essentially identical to conventional DFA
procedure (see \textit{e.g.}
\cite{Peng94,murad,physa,kunhu,kunhu1}). Suppose that $x_k$ is a
series of length $N$, and it is of compact support, \textit{i.e.}
$x_k = 0$ for an insignificant fraction of the values only.

\noindent $\bullet$ {\it Step 1}: Determine the profile
\begin{equation} Y(i) \equiv \sum_{k=1}^i \left[ x_k - \langle x
\rangle \right], \qquad i=1,\ldots,N. \label{profile}
\end{equation}
Subtraction of the mean $\langle x \rangle$ from $x_k$ is not
compulsory, since it would be eliminated by later detrending in
third step.

\noindent $\bullet$ {\it Step 2}: Divide profile $Y(i)$ into $N_s
\equiv {\rm int}(N/s)$ non-overlapping segments of equal lengths
$s$. Since length $N$ of series is often not a multiple of
considered time scale $s$, a short part at the end of profile may
remain. In order not to disregard this part of the series, same
procedure should be repeated starting from the opposite end.

\noindent $\bullet$ {\it Step 3}: Calculate  local trend for each of
$ N_s$ segments by a least-square fit of the series. Then determine
the variance
\begin{equation} F^2(s,\nu) \equiv {1 \over s} \sum_{i=1}^{s}
\left\{ Y[(\nu-1) s + i] - y_{\nu}(i) \right\}^2, \label{fsdef}
\end{equation}
for each segment $\nu = 1, \ldots, N_s$. Where, $y_{\nu}(i)$ is fitted
polynomial in segment $\nu$.  Linear, quadratic, cubic, or higher
order polynomials can be used in fitting procedure (conventionally
called DFA1, DFA2, $\ldots$, DFA$m$) \cite{Peng94,PRL00}.
In (MF-)DFA$m$, trend
of order $m$ in profile (and equivalently, order $m - 1$ in original
series) are eliminated. Thus a comparison of results for different
orders of DFA allows one to estimate the type of the polynomial
trend in the series \cite{physa,kunhu}.

\noindent $\bullet$ {\it Step 4}: Average over all segments to
obtain $q$-th order fluctuation function, defined by:
\begin{equation} F_q(s) \equiv \left\{ {1 \over N_s}
\sum_{\nu=1}^{N_s} \left[ F^2(s,\nu) \right]^{q/2} \right\}^{1/q},
\label{fdef}
\end{equation}
where, in general, variable $q$ can take any real value except zero.
For $q=2$, standard DFA procedure is retrieved. Generally we are
interested to know  how generalized $q$ dependent fluctuation
functions $F_q(s)$ depend on time scale $s$ for different values of
$q$. Hence, we must repeat steps 2, 3 and 4 for several scales
$s$. It is apparent that $F_q(s)$ will increase with increasing $s$.
Of course, $F_q(s)$ depends on  DFA order $m$. By construction,
$F_q(s)$ is only defined for $s \ge m+2$.

\noindent $\bullet$ {\it Step 5}: Determine scaling behavior of
fluctuation functions by analyzing log-log plots of $F_q(s)$ versus
$s$ for each value of $q$. If series $x_i$ are long-range power law
correlated, then $F_q(s)$, for large values of $s$, increases as a
power-law \textit{i.e.},
\begin{equation}
F_q(s) \sim s^{h(q)} \label{Hq}.
\end{equation}
In general, exponent $h(q)$ may depend on $q$. For stationary
series such as fractional Gaussian noise (fGn), $Y(i)$ in
Eq.~\ref{profile} will have a fractional Brownian motion (fBm)
signal, so, $0<h(q=2)<1.0$. The exponent $h(2)$ is identical to well
known Hurst exponent $H$ \cite{Peng94,murad,feder88}. Also, for
non-stationary signals, such as fBm noise, $Y(i)$ in
Eq.~\ref{profile} will be a sum of fBm signal, so corresponding
scaling exponent of $F_q(s)$ is identified by $h(q=2)>1.0$
\cite{Peng94,eke02}. For monofractal series, $h(q)$ is
independent of $q$, since scaling behavior of variance $F^2(s,\nu)$
is identical for all segments $\nu$, and averaging procedure in
Eq.~(\ref{fdef}) will just give a same scaling behavior for all
values of $q$. If we consider positive values of $q$, the segments
$\nu$ with large variance $F^2(s,\nu)$ (\textit{i.e.} large
deviation from the corresponding fit) will dominate average
$F_q(s)$. Thus, for positive values of $q$, $h(q)$ describes scaling
behavior of segments with large fluctuations. For negative values of
$q$, segments $\nu$ with small variance $F^2(s,\nu)$ will dominate
average $F_q(s)$. Hence, for negative values of $q$, $h(q)$
describes scaling behavior of segments with small fluctuations.

For a stationary and normalized series, multifractal scaling
exponent $h(q)$ defined in Eq.~(\ref{Hq}) is directly related to
scaling exponent $\tau(q)$ defined by standard partition function
based on multifractal formalism with the following analytical
relation (\textit{i.e.} see \cite{feder88})
\begin{equation}
\tau(q) = q h(q) - 1. \label{tauH}
\end{equation}
Thus, we see that $h(q)$ defined in Eq.~(\ref{Hq}) for MF-DFA is
directly related to classical multifractal scaling exponent
$\tau(q)$ and generalized
multifractal dimension \cite{feder88}
\begin{equation}
D(q) \equiv {\tau(q) \over q-1} =
{q h(q)-1 \over q-1}, \label{Dq}
\end{equation}
that is used instead of $\tau(q)$ in some papers. In this case,
while $h(q)$ is independent of $q$ for a monofractal series,
$D(q)$ depends on $q$. Another way to characterize a multifractal
series is looking to singularity spectrum $f(\alpha)$, which is
related to $\tau(q)$ via a Legendre transform \cite{feder88,peitgen}
\begin{equation}
\alpha = \tau'(q) \quad {\rm and} \quad
f(\alpha) = q \alpha - \tau(q). \label{Legendre}
\end{equation}
Here, $\alpha$ is singularity strength or H\"older exponent, while
$f(\alpha)$ denotes dimension of a subset of series that is
characterized by $\alpha$. Using Eq.~(\ref{tauH}), we can directly
relate $\alpha$ and  $f(\alpha)$ to $h(q)$
\begin{equation}
\alpha = h(q) + q h'(q) \quad {\rm and} \quad
f(\alpha) = q [\alpha - h(q)] + 1.\label{Legendre2}
\end{equation}
H\"older exponent denotes monofractality, while in multifractal
case, different parts of structure are characterized by different
values of $\alpha$, leading to existence of spectrum $f(\alpha)$.

\begin{figure}
\begin{tabular}{ccc}
\includegraphics[width=8cm]{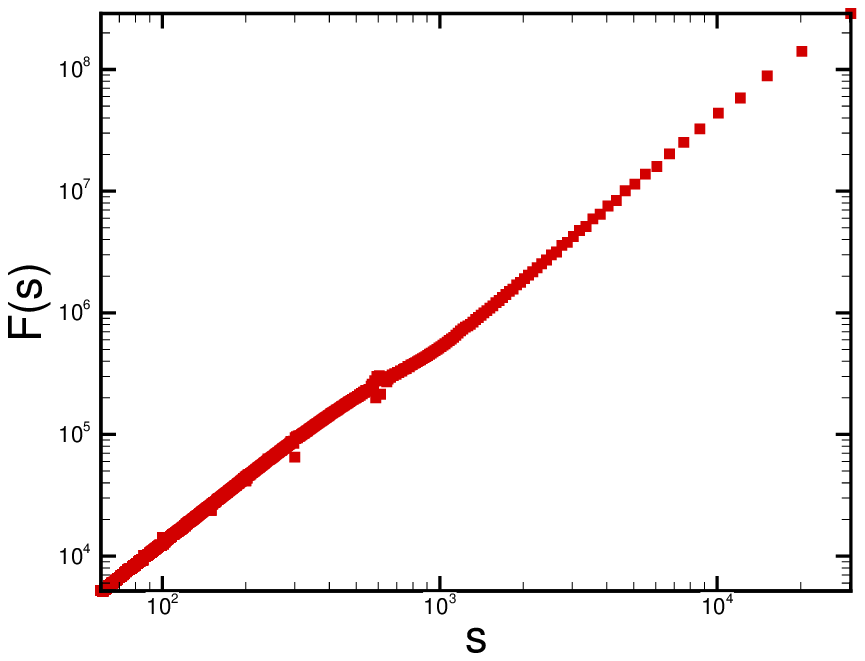}
 &\hspace{1.cm}&
\includegraphics[width=8cm]{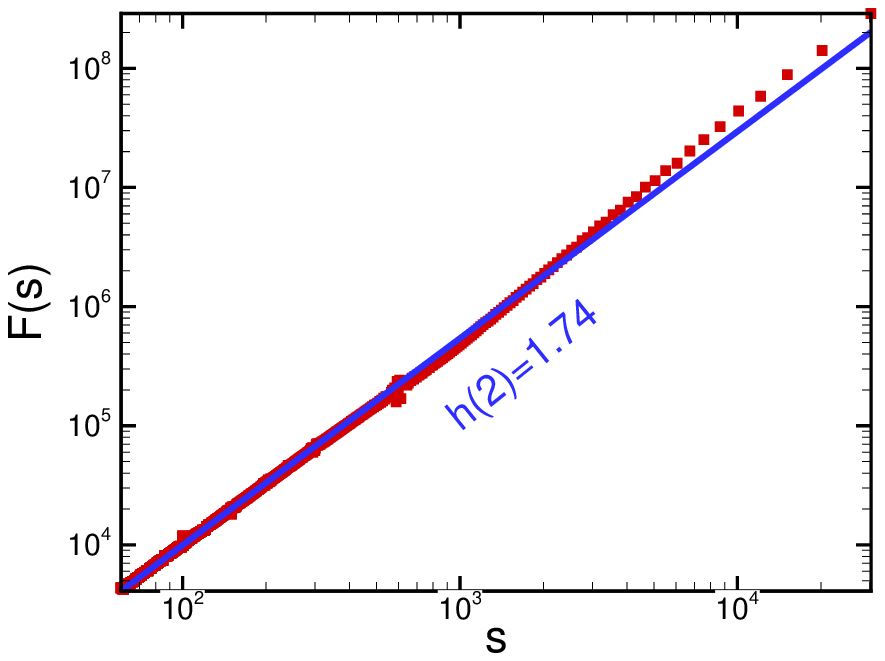}
\end{tabular}
\caption{ \label{fig4}The log-log plot $F(s)$ versus $s$ for $q=2.0$
after double profiling for gray style, before F-DFA (Left) and After
F-DFA (Right).}
\end{figure}

\begin{figure}
\centering
\includegraphics[width=9cm]{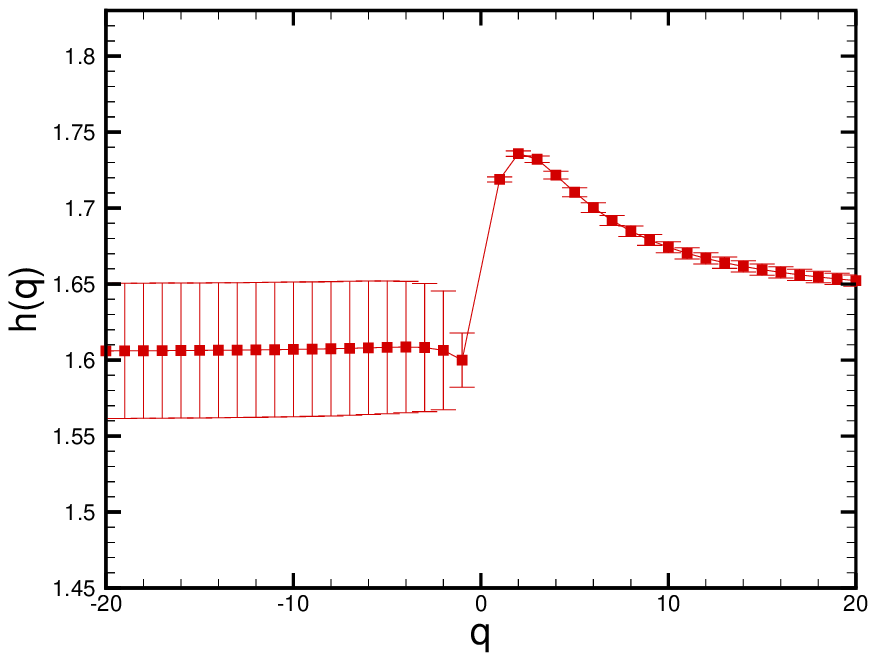}
\\
\includegraphics[width=9cm]{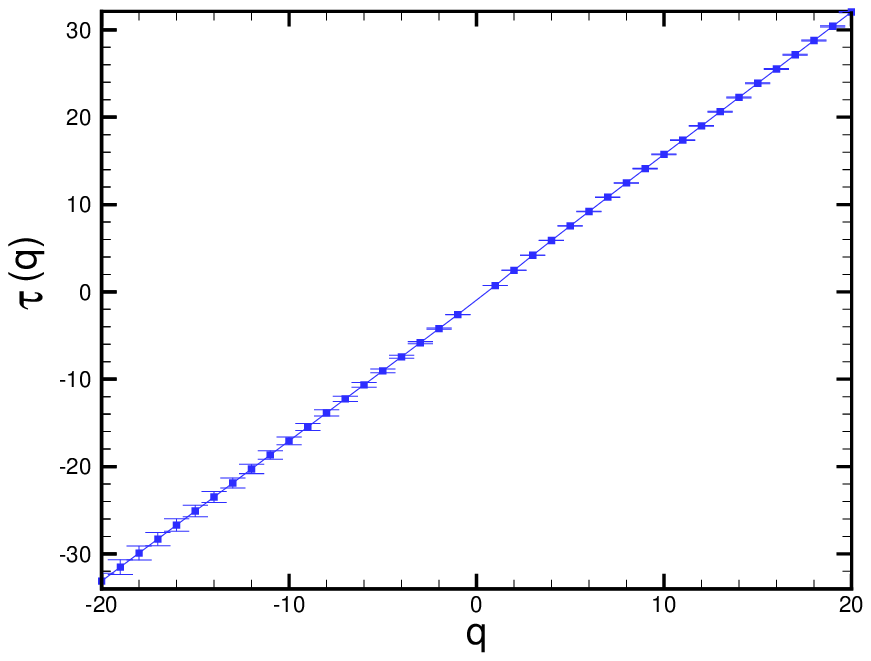}
\\
\includegraphics[width=9cm]{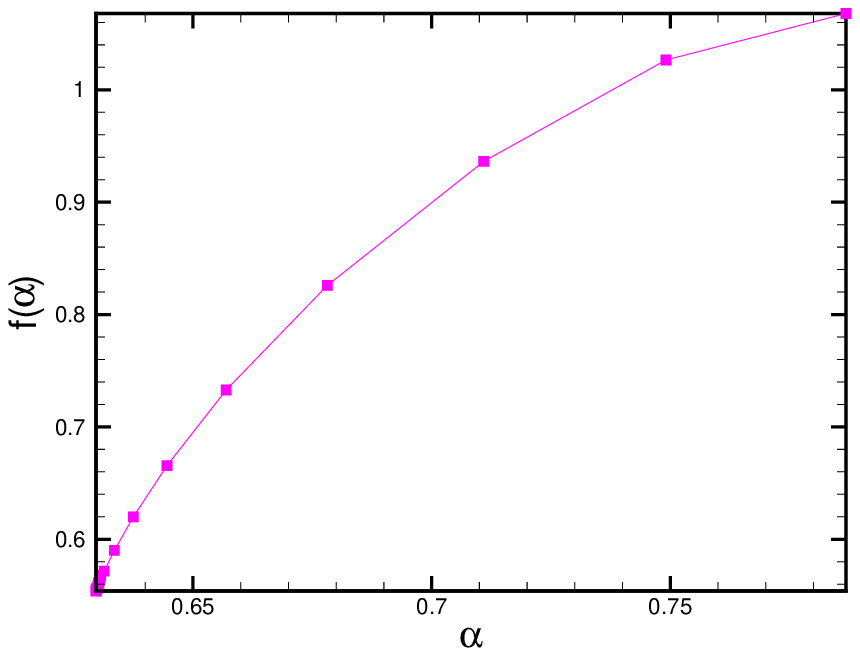}
\caption{\label{fig5}The q dependence of the exponents $h(q)$, $\tau(q)$
and singularity spectrum $f(\alpha)$, after double profiling, are
shown in the upper to lower panels respectively for gray style.}
\end{figure}

\subsection{Fourier-Detrended Fluctuation Analysis}
In some cases, there exists one or more crossover length scales
$s_l$ in graph $F(s)$, separating regimes with different scaling
exponents \cite{physa}. In these cases, obtaining the scaling
behavior is more complicated and different scaling exponents are
required for different parts of the series \cite{kunhu1}. Therefore,
one needs a multitude of scaling exponents in various scales for a
full description of scaling behavior. A crossover usually can arise
from a change in correlation properties of the signal at different
scales, or can often arise from trends in data. In addition,
inappropriate detrending can create additional trends to signal
which appear as crossovers in $F(s)$. The extra crossovers are not
actual length scales of the original signal. In fact, these
crossovers are due to the unsuitable detrending process.
Fourier-Detrend Fluctuation Analysis (F-DFA) can be applied to
remove crossovers such as sinusoidal trends. The F-DFA is a modified
approach for the analysis of low frequency trends
\cite{na04,chi05,koscielny98b}.

We transform data record to Fourier space in order to remove trends
having a low frequency periodic behavior. Then we truncate the first
few coefficients of the Fourier expansion and inverse Fourier
transform of the series. After removing the sinusoidal trends we can
obtain the fluctuation exponent using the direct calculation of the
MF-DFA. If truncation numbers are sufficient, The crossovers due to
sinusoidal trends in the log-log plot of $F_q(s)$ versus $s$
disappear.

\section{Analysis of Color series} \label{sec3}
Most paintings are non-stationary due to the fact that all
information is in whole of their structures. It means if we cut a
part of a painting, we can not complete the remainder of it.
Although there are some fractal paintings which all information is
in each part of them. As mentioned in Sec.~\ref{sec2}, a spurious of
correlations may be detected using detrending methods in
non-stationarity series. While, direct calculation of correlation
behavior, spectral density exponent, fractal dimensions, etc., do
not give us reliable results.

It seems that most of non-stationary properties in a painting are
known trends such as objects. In fact, a painting consists of some
objects or components have been set in various parts of it. As an
example, a portrait has different parts such as eyes, nose, mouth,
cheeks, eyebrows, etc. which can be considered as known trends.
Also, light and shadow can be considered as trends. Since the art of
painting is drawing the objects in an expressive manner, we intend
to detrend data to characterize the artistic work of a painter
regardless of the existence of specific objects (trends) in his or
her paintings.

To construct color series, we need to relate a standard numerical
quantity to each pixel. The basic colors are Red, Green and Blue
(RGB), which every color is a specific combination of these basic
colors. The standard value of each basic color is in the range of
$0-255$. Therefore, we consider each color in our analysis and also
we cover the gray style, which is a certain combination of these
three basic colors. In the first step, it can be checked out that,
all color series are non-stationary. One can verify non-stationarity
property experimentally by measuring stability of their average and
variance in a moving window for example using scale $s$. In fact,
there are some specific lengths in painting that have important
effects in statistical parameters. The picture's width and height
and color intensities are the first scaling parameters in painting.
Moreover, the body and face detail structures are important scales
that painter usually regards them. Let us determine whether the data
set has a sinusoidal trend or not. According to MF-DFA1 method,
Generalized Hurst exponents $h(q)$ in Eq.~(\ref{Hq}) can be found by
analyzing log-log plots of $F_q(s)$ versus $s$ for each $q$
(Fig.~\ref{fig4}). Our investigation shows that there is one
crossover length scale $s_{l}$ in log-log plots of $F_q(s)$ versus
$s$ for every $q$'s. To cancel the sinusoidal trends in MF-DFA1, we
apply F-DFA method for color data. For eliminating the crossover
scales, we need to remove only one term of the Fourier expansion.
Then, by inverse Fourier Transformation, the noise without
sinusoidal trend is extracted. Hurst exponent is between $0<H<1$.
However, MF-DFA method can only determine positive generalized Hurst
exponents, in order to refine the analysis near the fGn/fBm boundary
or strongly anti-correlated signals when it is close to zero.

The simplest way to analyze such data is to integrate series before
MF-DFA procedure. Hence, we replace the single summation in
Eq.~\ref{profile}, which is describing the profile from the original
data, by a double summation using signal summation conversion method
(SSC) \cite{eke02,Movahed,bun02,bach}. After using SSC method, fGn
switch to fBm and fBm switch to sum-fBm. In this case the relation
between the new exponent, $h(q=2)$, and $H$ is $H= h(q = 2)-1$
\cite{eke02,Movahed,bun02} (recently Movahed \textit{et al.} have
proven the relation between derived exponent from double profile of
series in DFA method and $h(q = 2)$ exponent in
Ref.~\cite{Movahed}). We find $h(q=2)=1.74 \pm 0.01$ for gray style
color series using SSC method. Therefore, Hurst exponent equals to
$H=h(q=2)-1=0.74$.

The results of MF-DFA1 method for color signal are shown in
Fig.~\ref{fig5}, which show that these color series can be
considered approximately as a monofractal process which is indicated
by weak $q$ dependence of the exponents $h(q = 2)$ and $\tau(q)$
\cite{bun02}. The $q$ dependence of multifractal scaling exponent
$\tau(q)$ approximately has a linear dependence to $q$ with equal
slopes as $1.63$, $1.59$, $1.70$, and $1.66$ for gray, red, green,
and blue styles, respectively. Table \ref{Tab} shows the obtained
quantities using MF-DFA1 method. Figure \ref{fig5} shows the width
of singularity spectrum ($f(\alpha)$) for gray style with $\Delta
\alpha\simeq0.15$ ($\Delta \alpha=\alpha(q_{max})-\alpha(q_{min})$).
Its value indicates that the power of multifractality of the color
series is weak \cite{paw05}.
\begin{table}[t]
\begin{center}
 \caption{\label{Tab}Values of $h(q=2)$, $\tau(q=2)$ exponents and
width of singularity spectrum, $f(\alpha)$, $\Delta \alpha$ for
$q=2.0$ of various styles obtained by MF-DFA1.}
%\medskip
\begin{tabular}{|c|@{\hspace{0.5cm}}c@{\hspace{0.5cm}}|@{\hspace{0.5cm}}c@{\hspace{0.5cm}}|@{\hspace{0.5cm}}c@{\hspace{0.5cm}}|}
 \hline
 Style    & $h(2)$         & $\tau(2)$     & $\Delta\alpha$   \\ \hline
 Gray     & $1.74\pm0.01$  & $2.48\pm0.01$ & $0.15$           \\ \hline
 Red      & $1.74\pm0.01$  & $2.48\pm0.01$ & $0.11$           \\ \hline
 Green    & $1.72\pm0.01$  & $2.44\pm0.01$ & $0.12$           \\ \hline
 Blue     & $1.74\pm0.01$  & $2.48\pm0.01$ & $0.12$           \\ \hline
\end{tabular}
\end{center}
\end{table}

\section{Conclusions}\label{sec4}
Using detrended methods for analyzing paintings, we can study the
artistic manner of the painter, regardless of subject of the
artworks. This is due to the fact that a painting consists of
objects which can be considered as trends. The non-stationary
property of paintings can be related to the objects, light and
shadow, etc. Indeed by detrending methods, we eliminate trends from
paintings and make the signal stationary. It means, we reduce the
effects of objects, light and shadow, etc. from paintings. We have
used multifractal detrended fluctuation analysis to detrend data to
characterize the most famous artistic work of Leonardo da Vinci and
shown the fractality of this painting. Our results using MF-DFA show
that color series of Mona Lisa almost has the same behavior in
various scales for all studied color styles (red, green, blue and
gray).
\section*{Acknowledgements}
We would like to thank M. Mirzaei and M. Vahabi for useful
discussion and comments.

\end{document}